# Frequency and Phase Synchronization in Neuromagnetic Cortical Responses to Flickering-Color Stimuli

**S.F. Timashev**[a,b,c], **Yu.S. Polyakov**[c], **R.M. Yulmetyev**[d,e], **S.A. Demin**[d,e], **O.Yu. Panischev**[d,e], **S. Shimojo**[f], **and J. Bhattacharya**[g,h]

[a] Institute of Laser and Information Technologies, Russian Academy of Sciences, ul. Pionerskaya, 2, Troitsk 142190, Russia

[b] Karpov Institute of Physical Chemistry, ul. Vorontsovo pole 10, Moscow 105064, Russia

[c] USPolyResearch, Ashland, PA 17921, USA

[d] Department of Physics, Kazan State University, ul. Kremlevskaya 18, Kazan 420008, Russia

[e] Department of Physics, Tatarstan State University of Humanities and Education, ul. Tatarstan 2, Kazan 420021, Russia

[f] Division of Biology, California Institute of Technology, MC 139-74, Pasadena, CA 91125, USA

[g] Department of Psychology, Goldsmiths College, University of London, New Cross, London SE14 6NW, UK

[h] Commission for Scientific Visualization, Austrian Academy of Sciences, Vienna, A1220, Austria

**ABSTRACT**

In our earlier study dealing with the analysis of neuromagnetic responses (magnetoencephalograms - MEG) to flickering-color stimuli for a group of control human subjects (9 volunteers) and a patient with photosensitive epilepsy (a 12-year old girl), it was shown that Flicker-Noise Spectroscopy (FNS) was able to identify specific differences in the responses of each organism. The high specificity of individual MEG responses manifested itself in the values of FNS parameters for both chaotic and resonant components of the original signal. The present study applies the FNS cross-correlation function to the analysis of correlations between the MEG responses simultaneously measured at spatially separated points of the human cortex processing the red-blue flickering color stimulus. It is shown that the cross-correlations for control (healthy) subjects are characterized by frequency and phase synchronization at different points of the cortex, with the dynamics of neuromagnetic responses being determined by the low-frequency processes that correspond to normal physiological rhythms. But for the patient, the frequency and phase synchronization breaks down, which is associated with the suppression of cortical regulatory functions when the flickering-color stimulus is applied, and higher frequencies start playing the dominating role. This suggests that the disruption of correlations in the MEG responses is the indicator of pathological changes leading to photosensitive epilepsy, which can be used for developing a method of diagnosing the disease based on the analysis with the FNS cross-correlation function.

## 1. INTRODUCTION

Frequency and phase synchronization, presence of the correlations between characteristic frequencies and phases of the excitations in different parts of the cortex (specific neural ensembles) when the actual signals are uncorrelated, and amplitude synchronization of the excitations are the necessary conditions for the brain to function as an integral system [1, 2]. Normally functioning brain reacts to external actions on the human organism by establishing some optimal level of such cross-correlations. A significant deviation from this optimal level, i.e., high level of synchronization or lack of synchronization, often points to a disease, such as epilepsy or tremor [3]. As the allowed level of synchronization is highly individual for each organism, the estimation of the normal level of such correlations for each organism by studying the recorded neural signals (i.e., the electroencephalograms (EEG) [3-5] or the magnetoencephalograms (MEG) [6-8]), is an important problem in modern medicine.

This study shows how the dynamic correlations between different parts of the cortex can be determined by the analysis of MEG signals with Flicker-Noise Spectroscopy (FNS) [9-12], a phenomenological approach to the analysis of any time series $V(t)$. The main difference of FNS compared to other related methods is the separation of the original signal into three frequency bands: system-specific “resonances” and their interferential contributions at lower frequencies, chaotic “random walk” (“irregularity-jump”) components at larger frequencies, and chaotic “irregularity-spike” (inertial) components in the highest frequency range. Specific parameters corresponding to each of the bands are introduced and calculated. These irregularities as well as specific resonance frequencies are considered as the information carriers on every hierarchical level of the evolution of a complex natural system with intermittent behavior, consecutive alternation of rapid chaotic changes in the values of dynamic variables on small time intervals with small variations of the values on longer time intervals (“laminar” phases) [13]. It was shown that the origins of intermittency are associated with the occurrence of complex (multiparticle, nonlinear) interactions, dissipation, and inertia in these systems [13].

The FNS approach was used in Ref. [14] for the analysis of the MEG signals recorded as the neuromagnetic response from a group of healthy human subjects (9 volunteers) and a patient (a 12-year girl) with photosensitive epilepsy while they were viewing equiluminant flickering stimuli of different color combinations (RB – red-blue and RG – red-green) [6, 7]. Photosensitive epilepsy (PSE) is the occurrence of high neural activity in response to visual stimuli, particularly flickering light, which is accompanied with various clinical and paraclinical manifestations. The interest to such analysis was triggered, in particular, by the perceived potential danger of modern cartoons to provoke PSE in children and adults. The experimental data were generated by 61 SQUID (superconducting quantum interference device) sensors attached to different points around the head, which can record weak magnetic induction gradients of about $10^{-11}$-$10^{-10}$ T/cm [6, 7]. The sampling frequency $f_d$ of MEG signals was 500 Hz ($f_d = 500\ Hz$).

The FNS analysis presented in Ref. [14] showed that there are several dynamic differences in the behavior of correlation links for the MEG signals in the patient compared to the corresponding signals in control subjects. It also confirmed our earlier conclusion [8] about the qualitative differences in power spectrum estimates for the dynamics of neuromagnetic cortex activity between the patient and the control subjects. One of the key differences is the occurrence of a high-frequency resonance near 100 Hz in the patient signals for sensor 10, located at the frontal lobe on the head; sensor 59, located at the central zone; and sensors 46, 51, and 53, located at the forehead.

In this paper, we extend our our results by analyzing the dynamic cross-correlations between spatially separated areas of the cortex that occur in response to flickering-color stimuli. The analysis of cross-correlations and the rate of their loss in the MEG signals was performed for different sensor pairs. Section 2 introduces the basic FNS relations, in particular, the cross-correlation expression for the analysis of dynamic correlations in the signals that are simultaneously measured at different points in space. The results of the cross-correlation calculations for the MEG signals are presented in section 3. Section 4 shows that the cross-correlations in MEG signals can be associated with frequency and phase synchronization in the integral system of the brain under external actions on the organism. The final section contains some concluding remarks on the connection between pathological changes in the PSE patient and suppression of regulatory functions in its cortex.

## 2. BASIC RELATIONS

In FNS, all the introduced information is related to one of the fundamental concepts of statistical physics, the autocorrelation function

$$\psi(\tau) = \langle V(t)V(t+\tau) \rangle , \qquad (1)$$

where $\tau$ is the time lag parameter. The angular brackets stand for the averaging over time interval $T$:

$$\langle (...) \rangle = \frac{1}{T} \int_{-T/2}^{T/2} (...)dt . \qquad (2)$$

To extract the information contained in $\psi(\tau)$ ($\langle V(t) \rangle = 0$ is assumed), the transforms, or "projections", of this function, are analyzed; specifically, the cosine transforms (“power spectrum” estimates) $S(f)$, where *f* is the frequency:

$$S(f) = \int_{-T/2}^{T/2} \langle V(t)V(t+t_1) \rangle \cos(2\pi f t_1)\, dt_1 \qquad (3)$$

and its difference moments (Kolmogorov transient structural functions) of the second order $\Phi^{(2)}(\tau)$:

$$\Phi^{(2)}(\tau) = \left\langle \left[ V(t) - V(t+\tau) \right]^2 \right\rangle . \qquad (4)$$

The information contents of $S(f)$ and $\Phi^{(2)}(\tau)$ are different, and the parameters for both functions are needed to solve parameterization problems. By considering the intermittent character of signals under study, interpolation expressions for the chaotic components $\Phi_c^{(2)}(\tau)$ and $S_c(f)$ of $S(f)$ and $\Phi^{(2)}(\tau)$, respectively, were derived using the theory of generalized functions in Ref. [11]. It was shown that structural functions $\Phi^{(2)}(\tau)$ are formed only by jump irregularities, and functions $S(f)$, which characterize the "energy side" of the process, are formed by both types of irregularities, spikes and jumps.

In this regard, let us mention a well-known result presented in the Chapter 4.3 of Ref. [15] (see Fig. 51), where an intermittent chaotic signal with alternating rapid chaotic spikes and laminar phases was considered. An artificial signal was generated by introducing a sequence of Dirac delta functions instead of rapid chaotic spikes. Then the power spectral density $S_c(f)$ for a sequence of $\delta$-functions with characteristic time intervals $T_0^i$ between adjacent $\delta$-functions on macroscopic time intervals [$-T/2$, $+\ T/2$] ($T_0^i \ll T$) was calculated. It was shown that this artificial signal formed a flicker-noise dependency $S_c(f) \sim f^{-n}$ ($n \sim 1$) in the low-frequency spectrum range ($f \ll 1/2\pi T_0^i$). In other words, it was *informative*. On the other hand, if one would calculate the difference moment $\Phi_c^{(2)}(\tau)$ for this artificial signal (difference moment was not considered in [15]), it would be clear that it is equal to zero because the domain set of a $\delta$-function sequence is a set of measure zero [16]. It is easy to numerically illustrate this statement by replacing $\delta$-functions in calculating $\Phi_c^{(2)}(\tau)$ with one of the well-known approximations; for example, Gaussian with dispersion $\sigma_G^2$, and then passing to the limit $\sigma_G \to 0$. It should be underlined that such separation of information stored in various irregularities is attributed to the intermittent character of the evolution dynamics. Indeed, the information contents of $S_c(f)$ and $\Phi_c^{(2)}(\tau)$ coincide if there is no intermittence, as shown for the case of completely "irregular" dynamics of the Weierstrass – Mandelbrot (WM) function (see [17]).

The interpolation expressions $\Phi_c^{(2)}(\tau)$, $S_c(f)$, and the FNS parameterization algorithm are described in detail elsewhere [9, 10, 17]. The FNS parameters were calculated by fitting the chaotic interpolation expressions to the corresponding difference moments and power spectrum estimates for the experimental data. The introduced parameters are: $\sigma$, the standard deviation of the measured dynamic variable with dimension [$V$]; $H_1$, the Hurst constant, which describes the rate at which the dynamic variable "forgets" its values on the time intervals that are less than the correlation time $T_1$ (in this case, $T_1$ may be interpreted as the correlation time for the jumps in the stochastically varying time series $V(t)$); $S_c(0)$, the low-frequency limit of $S_c(f)$; and $n_0$, the degree of correlation loss in the sequence of spikes on time interval $T_0$.

FNS also includes cross-correlation expressions, which allow one to analyze different (mass, electric, magnetic) flows in distributed systems. The information about the dynamics of correlation links in variables $V_i(t)$ and $V_j(t)$, measured at different points $i$ and $j$, can be extracted by analyzing the temporal links of various correlators. Here, we will limit our attention to the simplest "two-point" correlation expression characterizing the links between $V_i(t)$ and $V_j(t)$ [9-11, 18]:

$$q_{ij}(\tau;\theta_{ij}) = \left\langle \left[ \frac{V_i(t) - V_i(t+\tau)}{\sqrt{\Phi_i^{(2)}(\tau)}} \right] \left[ \frac{V_j(t+\theta_{ij}) - V_j(t+\theta_{ij}+\tau)}{\sqrt{\Phi_j^{(2)}(\tau)}} \right] \right\rangle_{T-\tau-|\theta_{ij}|}, \quad (5)$$

where $\tau$ is the "lag time"; $\theta_{ij}$ is the "time shift" parameter.

The dependence of cross-correlation $q_{ij}(\tau;\ \theta_{ij})$ on $\theta_{ij}$ describes the cause-and-effect relation ("flow direction") between signals $V_i(t)$ and $V_j(t)$. When $\theta_{ij} > 0$, the flow moves from point $i$ to point $j$, when $\theta_{ij} < 0$, from $j$ to $i$. When the distance between points $i$ and $j$ is fixed, the value of $\theta_{ij}$ can be used to estimate the rate of information transfer between these two points. The dependence of the value and magnitude of cross-correlation $q_{ij}(\tau;\ \theta_{ij})$ on $\tau$ and $\theta_{ij}$ can be used to analyze the flow dynamics with signals $V_i(t)$ and $V_j(t)$ changing in phase ($q_{ij} > 0$) and in antiphase ($q_{ij} < 0$).

Just like in difference moments (4), there is no contribution by the highest-frequency chaotic spike component to cross-correlations (5) containing the averaged product of differences with time lag $\tau$. Therefore, the cross-correlation values are fully determined by the correlations in low-frequency "resonance" and irregularity-jump components of signals $V_i(t)$ and $V_j(t)$ recorded at points $i$ and $j$.

## 3. CROSS-CORRELATIONS IN NEUROMAGNETIC RESPONSES

Out of 1830 possible pair-wise combinations for cross-correlations in the MEG signals recorded by 61 SQUID sensors, consider the cross-correlations for sensors 10 and 59, which show the highest sensitivity to the color stimuli, and sensor 54, which is characterized by highly reduced anomalous manifestations in the PSE patient. In this study we will limit our attention to the analysis of the data for the PSE patient, 1$^{st}$ control subject, and 6$^{th}$ control subject. The 6$^{th}$ control subject was selected because its MEG responses are characterized by the least number of resonances, which made it easier to identify the key differences between the patient and healthy control subjects. All other control subjects had a very similar behavior of cross-correlations. The 1$^{st}$ control subject was chosen because it was the first control subject with this typical behavior.

Figures 1-3 illustrate 1.7-second segments of MEG signals recorded by sensors 10, 54, and 59 for RB stimulus. The MEG signals at all sensors for control subjects are characterized by large-scale fluctuations and quasi-oscillatory behavior while the signals for the PSE patient are characterized by small-scale fluctuations on a quasi-periodic background.

Figures 4-6 illustrate the low-frequency part of spectral estimate $S(f)$ and the difference moment $\Phi^{(2)}(\tau)$ calculated at sensors 10, 54, and 59 for the 1$^{st}$ control subject (Fig. 4), 6$^{th}$ control subject (Fig. 5), and the PSE patient (Fig. 6). These variations demonstrate that the MEG responses of each control subject are individual-specific. For example, the structure of power spectral estimates for the 6$^{th}$ control subject is much simpler than for the 1$^{st}$ control subject: it has a lesser number of characteristic frequencies. The structure of power spectral estimates for the PSE patient is much different from the spectral estimates for both control subjects.

The variations $S(f)$ in control subjects suggest that the key role in the dynamics of the appropriate signals is played by low-frequency processes (Figs. 5 and 6). The peak values of $S(f)$ at the frequencies of 9 and 18 Hz can be associated with the normal physiological rhythms that reflect complex psycho-physiological processes in brain activity. Changes in the rhythms imply some disruptions in normal brain functioning and the central nervous system of the organism. In addition to the peaks at characteristic frequencies, the spectral estimates also contain peaks reflecting other periodic features in bioelectric brain activity.

In the patient, however, the characteristic frequencies are seen, but to a lesser degree. At the same time, the key role in this case is played by the processes that have higher frequencies: 50 and 100 Hz (the only exception is the signal at "non-anomalous" sensor 54, which does not contain any peaks in the range higher than 20 Hz). Additional peaks characterizing the periodic processes in the neuromagnetic brain activity of the patient are also noticeable throughout the range of low frequencies (sensors 10 and 59). Anomalous periodic behavior in the signals for the patient, which can be attributed to physiological rhythms and other processes, suggests that there are substantial changes in the neuromagnetic brain activity of the patient compared to the control group.

This implies the cerebral cortex of a healthy (with respect to PSE) subject has some sort of protective mechanism that coordinates the frequency and phase synchronization and prevents the collective anomalous neural activity, which apparently leads to epileptic seizures, when dangerous flicker-color stimuli are applied. Pathological changes in the PSE patient suppress the regulatory activity of this natural controller, which is reflected in the high-frequency dynamics of brain signals.

It should be noted that the sum of resonance and chaotic FNS interpolations for $\Phi^{(2)}(\tau)$ in Figs. 4(b, d, f)-6(b, d, f) is in good agreement with the experimental difference moment determined using the source signal. The only exception is the data for sensor 59 of the patient shown in Fig. 6f, where some noticeable deviations can be seen. The corresponding values of the FNS parameters for MEG responses are listed in the figure captions. As can be seen from the analysis of the variations in Figs. 4(a, c, e)-6(a, c, e), the key information about normal and abnormal behavior in the MEG responses is contained in the low-frequency resonances of spectral estimates $S(f)$. The same information, which is highly individual, is also clearly seen in the curves for difference moments $\Phi^{(2)}(\tau)$. The FNS parameters for the chaotic components, which are also highly individual in this case, contain very little information useful for the characterization of normal and abnormal behavior.

Figures 7-9 illustrate the cross-correlations for sensor pairs 10-59, 10-54, and 54-59 for the 1$^{st}$ control subject (Figs. 7a-9a), 6$^{th}$ control subject (Figs. 7b-9b), and PSE patient (Figs. 7c-9c). Cross-correlation variations $q_{ij}$ ($\tau$; $\theta_{ij}$) for the control subjects are characterized by clear large-scale structures suggesting that the key role in the dynamics of the signals is played by low-frequency resonances, which can be seen from the analysis of the corresponding spectral estimates $S(f)$. As was noted above, the low-frequency part of $S(f)$ for the 1$^{st}$ control subject is characterized by a higher number of frequencies as compared to the 6$^{th}$ control subject. This leads to a more complex behavior in the cross-correlations for the 1$^{st}$ control subject (Figs. 7a-9a). On the other hand, the cross-correlations for the patient show small-scale oscillations, especially in the case of sensor pair 10-59, which reflect the dominating role played by the processes with the frequencies that are higher than for the control subjects.

## 4. FREQUENCY AND PHASE SYNCHRONIZATION IN THE NEUROMAGNETIC ACTIVITY REGISTERED IN MEG RESPONSES

Consider the cross-correlations $q_{ij}$ ($\tau$; $\theta_{ij}$) for MEG responses in the 6$^{th}$ control subject, where the periodic behavior is most pronounced. Let us look at the variation for sensor pair 10-59 (Fig. 7b) and one of its cross sections $q_{10\text{-}59}$ ($\tau^0$; $\theta_{10\text{-}59}$) at $\tau^0 = 0.1T$ (Fig. 10). The closest to $\theta_{10\text{-}59} = 0$ values of $\theta_{10\text{-}59} = \theta_{10\text{-}59}^{0+}$ and $\theta_{10\text{-}59} = \theta_{10\text{-}59}^{0-}$ corresponding to the maximum values of correlation ($q_{10\text{-}59} \sim 0.5$) and anticorrelation ($q_{10\text{-}59} \sim -0.5$) are -56, -2, 51 $f_d^{-1}$ and –33 and 25 $f_d^{-1}$, respectively. The graphs of $q_{10\text{-}59}$ ($\tau$; $\theta_{10\text{-}59}^{0+}$) and $q_{10\text{-}59}$ ($\tau$; $\theta_{10\text{-}59}^{0-}$) for these values of $\theta_{10\text{-}59}^{0+}$ and $\theta_{10\text{-}59}^{0-}$ are illustrated in Figs. 11 and 12, respectively.

It is easy to see from Figs. 11 and 12 that $f \approx 10$ *Hz* is the determining frequency, which also clearly manifests itself in the spectral estimates $S(f)$ for sensors 11 (Fig. 5a) and 59 (Fig. 5e) of the 6-th control subject. The same frequency determines the periodic dependence of $q_{10\text{-}59}(\tau^0;\ \theta_{10\text{-}59})$ on $\theta_{10\text{-}59}$ with a period of $T_{10\text{-}59} \approx 50 f_d^{-1} \approx 0.1$ s (Fig. 10). It should also be noted that the curve in Fig. 10 is symmetric with respect to plane $\theta_{10\text{-}59}^{0+} = -2 f_d^{-1} \approx 0.004$ s.

According to Eq. (5), the fact that the maximum value of cross-correlation $q_{10\text{-}59}(\tau^0;\ \theta_{10\text{-}59})$ is reached at $\theta_{10\text{-}59}^{0+} = -56 f_d^{-1}$ implies that the response at sensor 59 precedes the response at sensor 10 by $T_a \approx 54 f_d^{-1} \approx 0.1$ s. At the same time, the occurrence of another comparable local maximum in $q_{10\text{-}59}(\tau^0;\ \theta_{10\text{-}59})$ at $\theta_{10\text{-}59}^{0+} = 51 f_d^{-1}$ implies that the response at sensor 59 happens later than the response at sensor 10 by $T_r \approx 53 f_d^{-1} \approx 0.1$ s. The same results apply to the analysis of the minimum values $\theta_{10\text{-}59}^{0-}$ corresponding to the highest anticorrelations for sensor pair 10-59. In other words, the presented cross-correlations for the 6-th patient demonstrate that the periodic changes at sensor 10 bring about the corresponding periodic changes at sensor 59, and vice versa. This conclusion, which appears to be contradictory at first glance, actually indicates that mutual synchronization takes places between the responses at sensors 10 and 59. In other words, there is a frequency and phase synchronization driven by the cortex system, which coordinates the responses to color-flickering stimuli at different parts of the cortex.

Same conclusions can be drawn from the analysis of cross-correlations for other sensor pairs of the 1st and 6th control subjects (Figs. 7-9). However, the frequency and phase synchronization in the 1st subject is more complicated than in the 6th one, which is attributed to a more complex behavior of low-frequency *S*(*f*) for the 1st control subject.

The behavior of cross-correlations in the MEG responses of the patient (Figs. 7c – 9c) is much different from the behavior in the control subjects (Figs. 7a,b – 9a,b). First, there is no symmetry with respect to plane $\theta_{ij}^{0+} \approx 0$. Second, the cross-correlations for the patient, especially in the case of the "anomalous" pair 10-59, are characterized by high-frequency oscillations at frequencies 50 and 100 Hz. This can be easily seen in Figs. 13 and 14 illustrating the cross-sections of variation $q_{10\text{-}59}(\tau;\ \theta_{10\text{-}59})$ (Fig. 7c) for sensor pair 10-59 at fixed values of $\tau^0$ and $\theta_{10\text{-}59}^{0+}$.

The characteristic time interval $\Delta t_p$ separating the local maximum (minimum) values near $\theta_{10\text{-}59} = 0$ for graph in Fig. 13 is $\Delta t_p \sim 10 f_d^{-1} \approx 0.02$ s, i.e., $f \approx 50$Hz, which is also seen in the spectral estimates *S*(*f*) for sensors 10 and 59 (Figs. 6a,e). The variation in Fig. 14 is more chaotic, with noticeable high-frequency components in the range up to 200 Hz. At the same time, the cross-correlations for sensor pair 54-59 also manifest significant low-frequency components, which are present in the spectral estimate for sensor 54 (Fig. 6c). Just like in Fig. 13, in this case there is no symmetry of $q_{54\text{-}59}(\tau;\ \theta_{54\text{-}59})$ with respect to plane $\theta_{54\text{-}59} \approx 0$.

The absence of the symmetry and low level of cross-correlations in the MEG responses of the patient can be regarded as a medical sign in the diagnosis of photosensitive epilepsy, which implies the total disruption of frequency and phase synchronization in the neuromagnetic responses of subject's cortex to the RB flickering-color stimulus. As additional analysis showed, this loss of symmetry manifests itself for other sensors, not just the "anomalous" sensor set of (10, 46, 51, 53, 59). Similar conclusions on the disruption of the synchronization in MEG responses to flickering-color stimuli were made in Refs. [6] and [7].

## 5. CONCLUDING REMARKS

The above FNS analysis of the neuromagnetic cortical responses to the RB flickering-color stimulus for two control human subjects and a PSE patient shows that the cortical dynamics in the controls is predominantly determined by the low-frequency processes that correspond to normal physiological rhythms. The pathological changes in the neuromagnetic activity of the PSE patient bring about the relaxation processes with higher frequencies and lesser degree of cross-correlations as compared to the responses in the control subjects.

It is shown that the disruption of the cross-correlations between different cortical regions, which leads to the suppression of cortical regulatory functions (frequency and phase synchronization) when external actions exceeding a certain threshold are applied, is the main indicator of the pathological changes taking place in the PSE patient.

The demonstrated effectiveness of FNS analysis in identifying the individual features of the MEG responses of not only a patient but also healthy subjects implies that the FNS approach may be used for finding subtle individual differences in the human cerebral activity resulting from external stimuli.

## ACKNOWLEDGMENTS

This study was supported in part by the Russian Foundation for Basic Research, projects no. 08-02-00230-*a* and 08-02-00123-*a*.

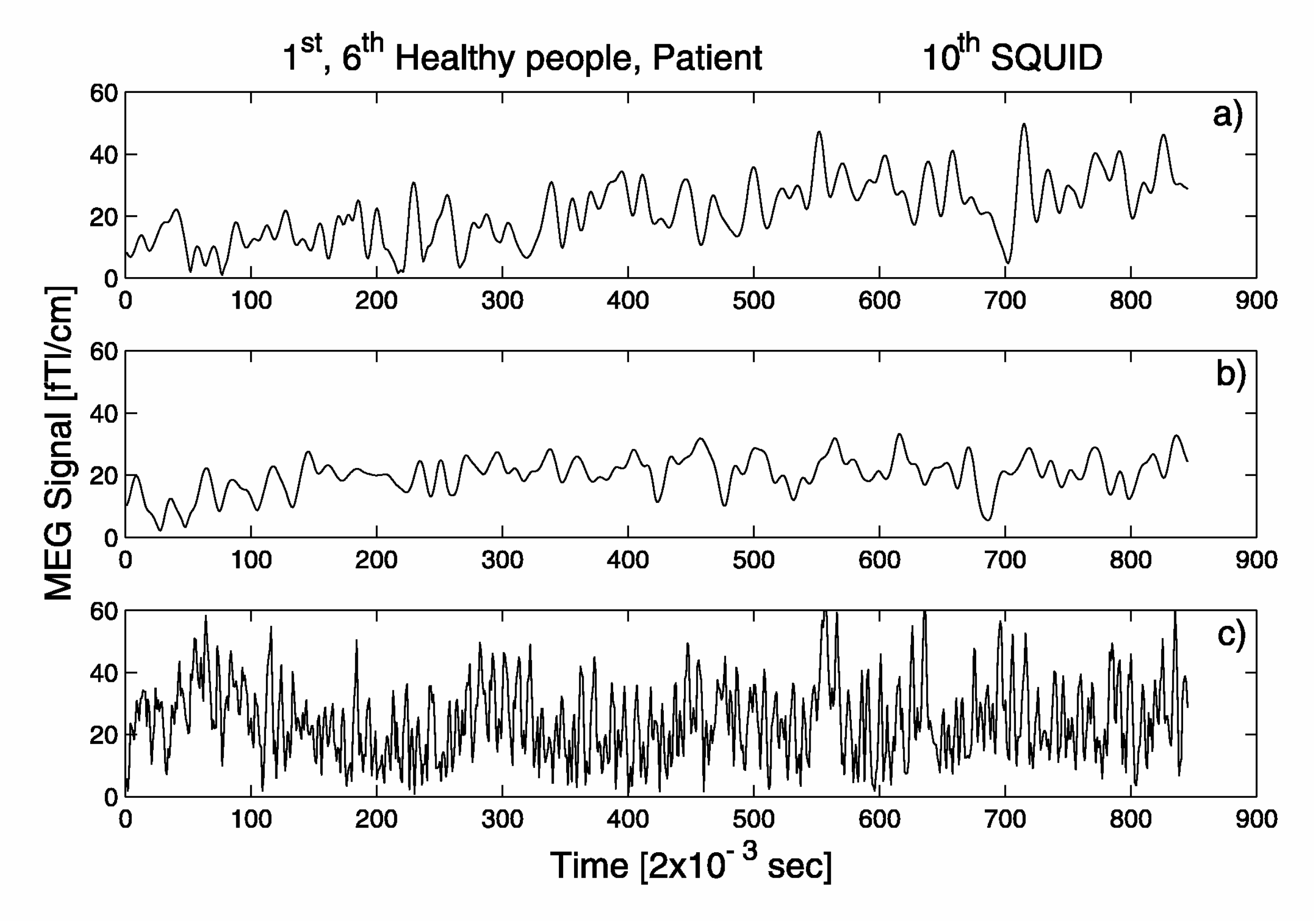

Fig. 1. MEG signals recorded at sensor 10 as the response to RB-stimulus for the 1st control subject (a), 6th control subject (b), and PSE patient (c).

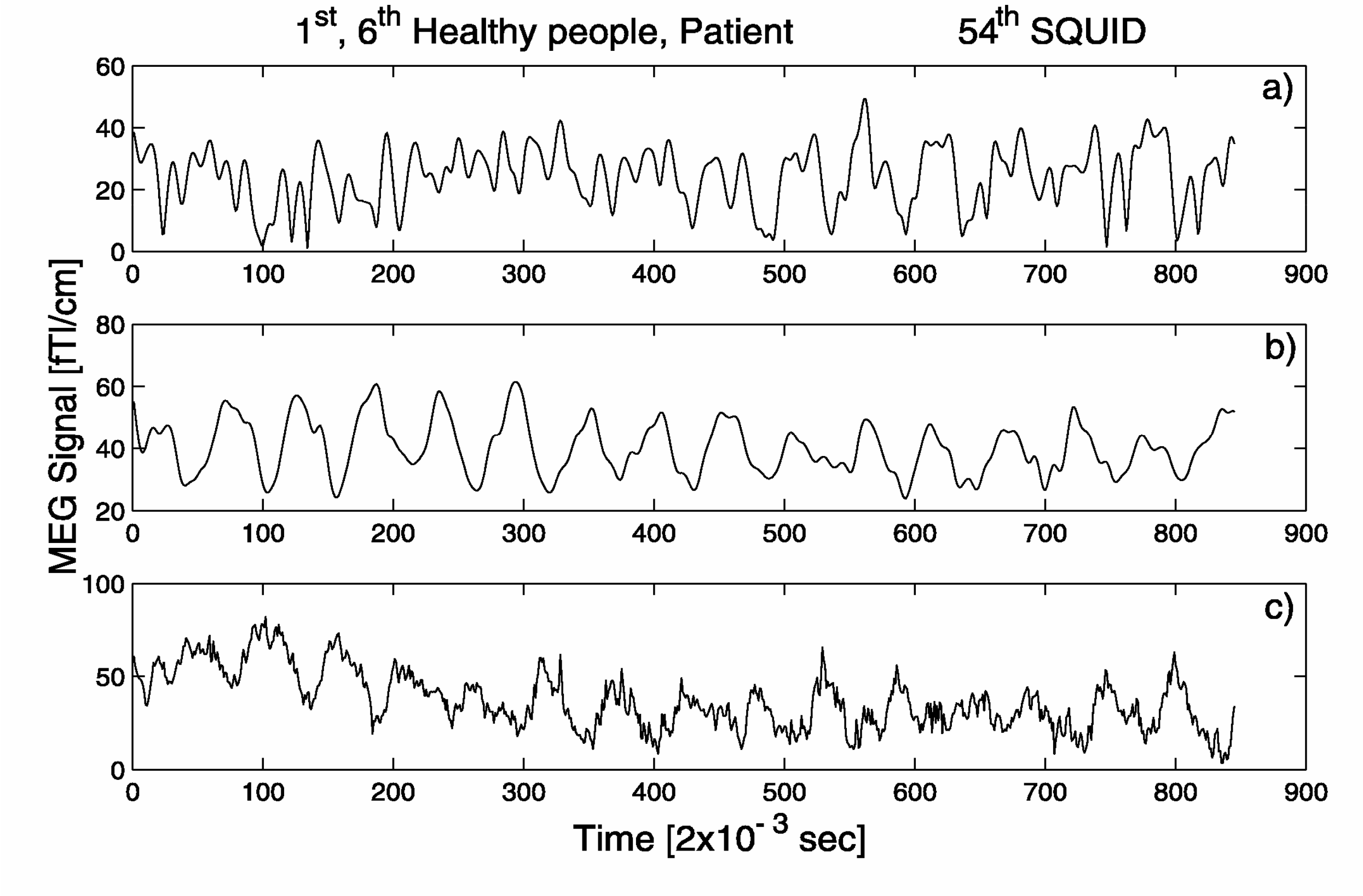

Fig. 2. MEG signals recorded at sensor 54 as the response to RB-stimulus for the 1st control subject (a), 6th control subject (b), and PSE patient (c).

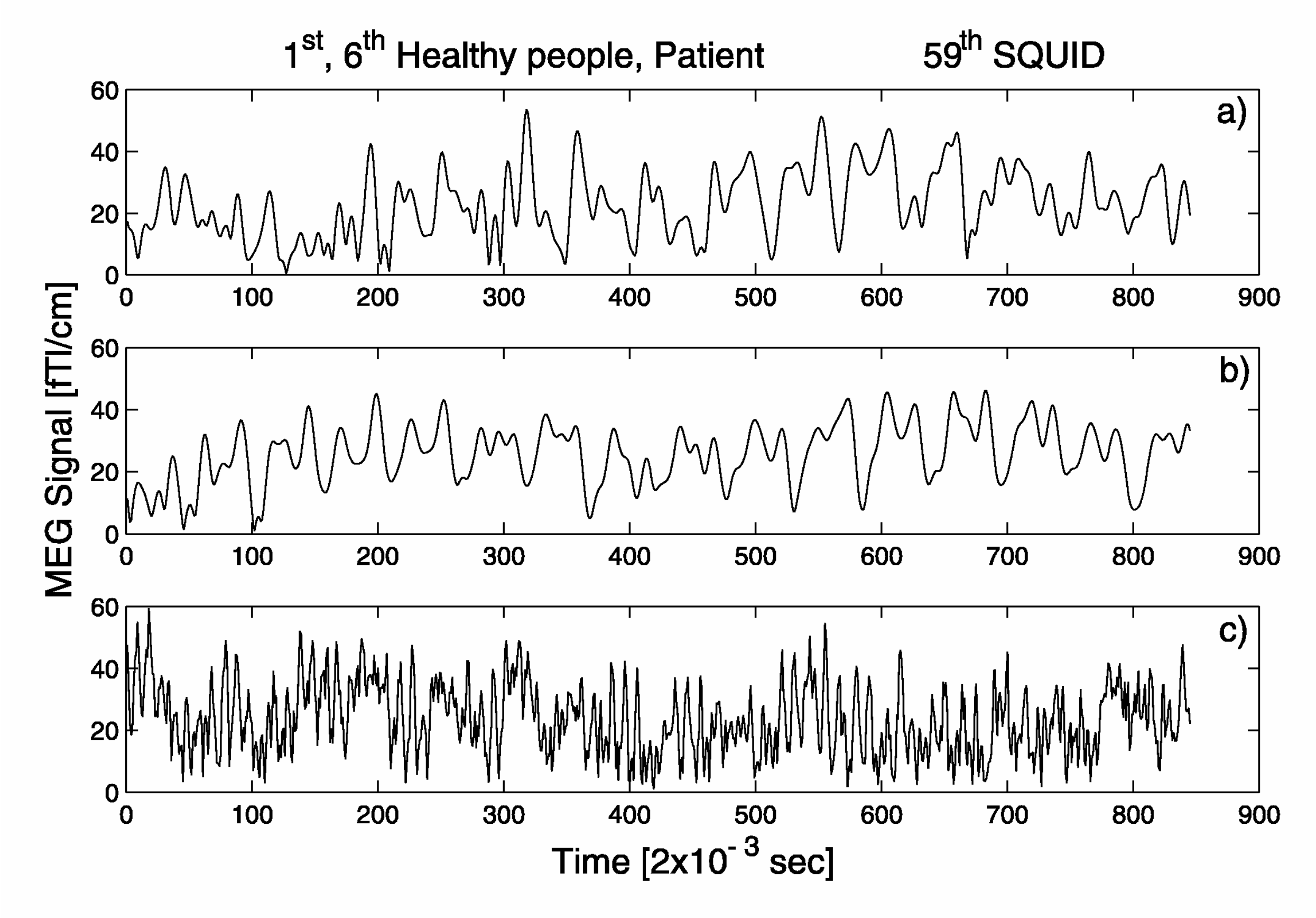

Fig. 3. MEG signals recorded at sensor 59 as the response to RB-stimulus for the 1st control subject (a), 6th control subject (b), and PSE patient (c).

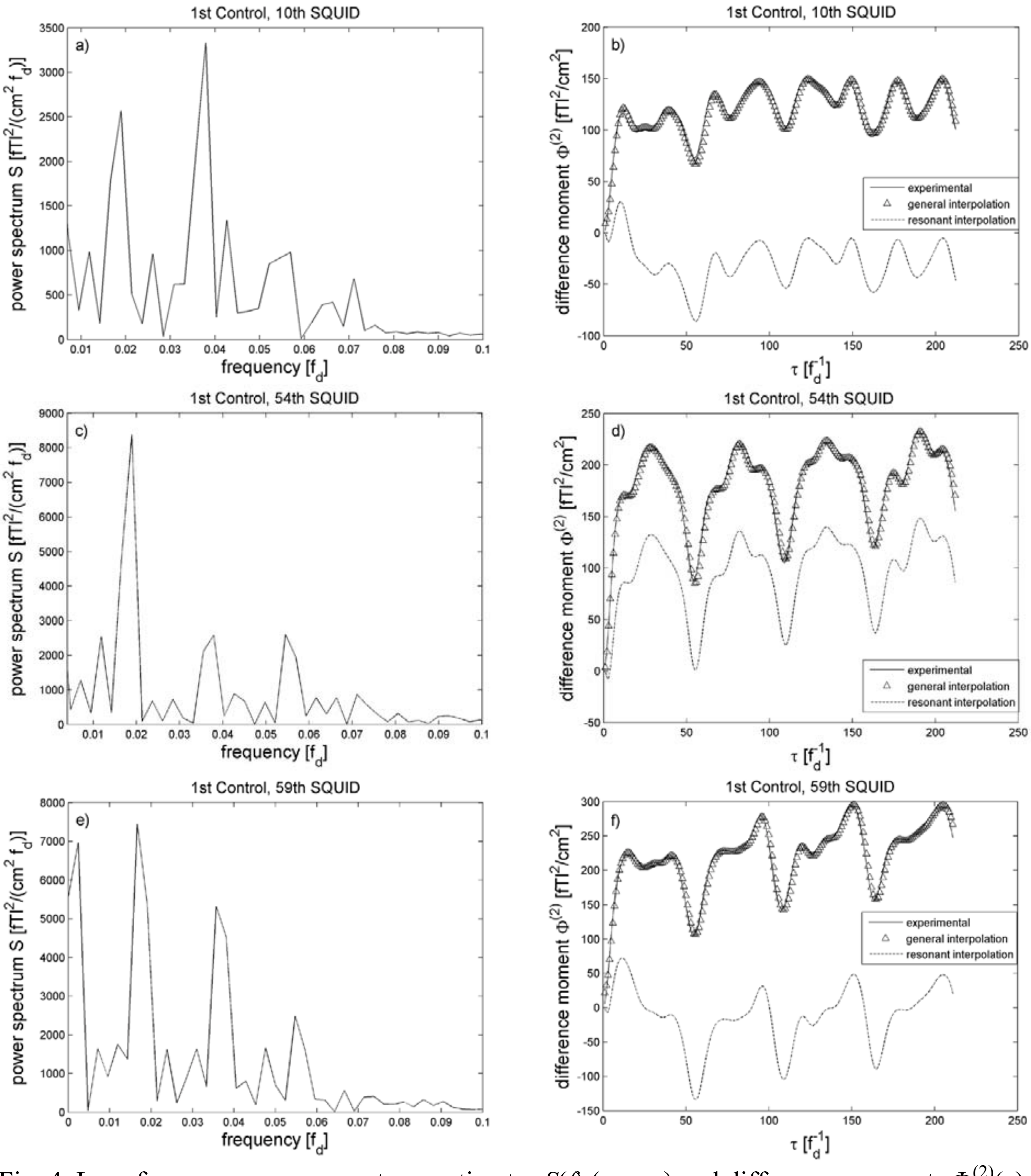

Fig. 4. Low-frequency power spectrum estimates $S(f)$ (a, c, e) and difference moments $\Phi^{(2)}(\tau)$ (b, d, f) for MEG responses at sensors 10, 54, and 59, respectively, in the 1st control subject. Values of FNS parameters: 10th sensor: $\sigma = 8.11$ fTl/cm, $S_c(0) = 4059.87$ $\mathrm{fT}^2 \cdot f_d^{-1}/\mathrm{cm}^2$, $H_1 = 0.47$, $T_1 = 16.49\, f_d^{-1}$, $T_0 = 14.67\, f_d^{-1}$, $n = 1.92$; 54th sensor: $\sigma = 6.47$ fTl/cm, $S_c(0) = 663.56$ $\mathrm{fT}^2 \cdot f_d^{-1}/\mathrm{cm}^2$, $H_1 = 2.42$, $T_1 = 0.88\, f_d^{-1}$, $T_0 = 3.33\, f_d^{-1}$, $n = 2.33$; 59th sensor: $\sigma = 11.17$ fTl/cm, $S_c(0) = 5521.92$ $\mathrm{fT}^2 \cdot f_d^{-1}/\mathrm{cm}^2$, $H_1 = 0.42$, $T_1 = 19.76\, f_d^{-1}$, $T_0 = 11.36\, f_d^{-1}$, $n = 2.03$.

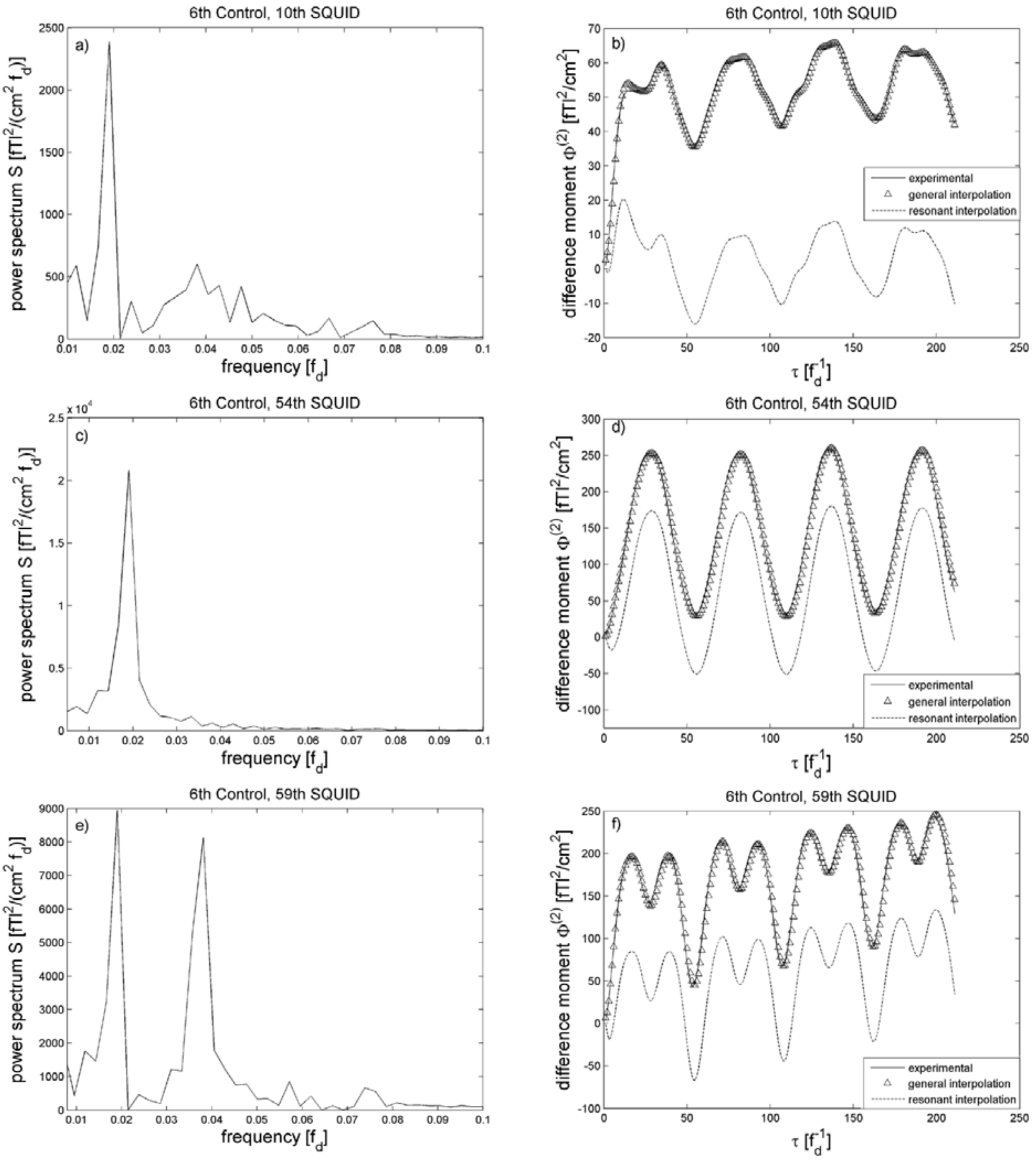

Fig. 5. Low-frequency power spectrum estimates $S(f)$ (a, c, e) and difference moments $\Phi^{(2)}(\tau)$ (b, d, f) for MEG responses at sensors 10, 54, and 59, respectively, in the $6^{th}$ control subject. Values of FNS parameters: $10^{th}$ sensor: $\sigma = 5.33$ fTl/cm, $S_c(0) = 1687.80$ fT$^2 \cdot f_d^{-1}$/cm$^2$, $H_1 = 0.83$, $T_1 = 9.27 f_d^{-1}$, $T_0 = 10.59 f_d^{-1}$, $n = 2.53$; $54^{th}$ sensor: $\sigma = 6.47$ fTl/cm, $S_c(0) = 663.56$ fT$^2 \cdot f_d^{-1}$/cm$^2$, $H_1 = 2.42$, $T_1 = 0.88 f_d^{-1}$, $T_0 = 3.33 f_d^{-1}$, $n = 2.33$; $59^{th}$ sensor: $\sigma = 10.75$ fTl/cm, $S_c(0) = 5852.75$ fT$^2 \cdot f_d^{-1}$/cm$^2$, $H_1 = 0.44$, $T_1 = 22.35 f_d^{-1}$, $T_0 = 15.19 f_d^{-1}$, $n = 1.86$.

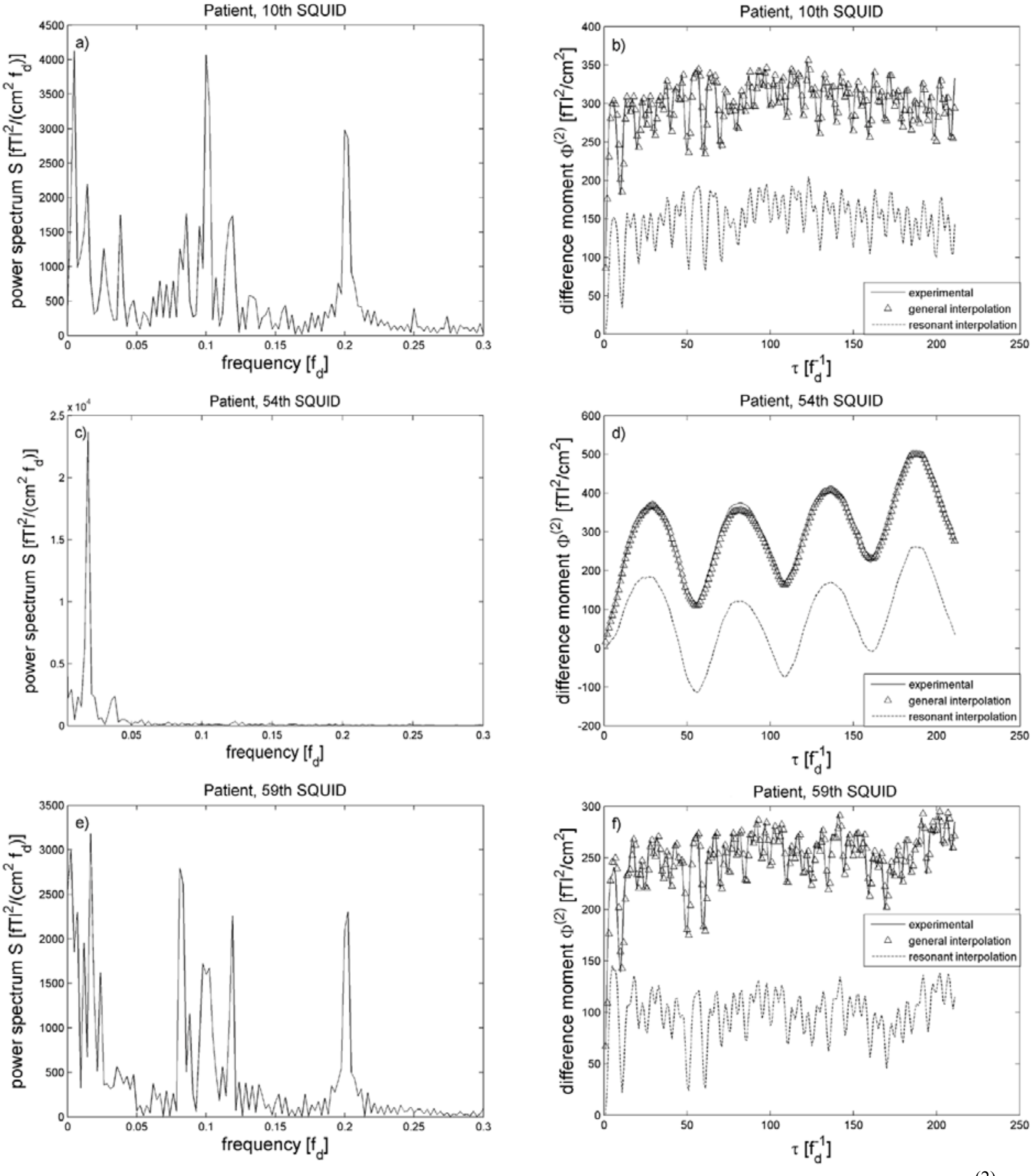

Fig. 6. Low-frequency power spectrum estimates $S(f)$ (a, c, e) and difference moments $\Phi^{(2)}(\tau)$ (b, d, f) for MEG responses at sensors 10, 54, and 59, respectively, in the PSE patient. Values of FNS parameters: 10th sensor: $\sigma = 8.28$ fTl/cm, $S_c(0) = 378.65$ fT$^2 \cdot f_d^{-1}$/cm$^2$, $H_1 = 7.36$, $T_1 = 0.11\, f_d^{-1}$, $T_0 = 1.03\, f_d^{-1}$, $n = 3.03$; 54th sensor: $\sigma = 9.30$ fTl/cm, $S_c(0) = 23865.20$ fT$^2 \cdot f_d^{-1}$/cm$^2$, $H_1 = 0.36$, $T_1 = 25.68\, f_d^{-1}$, $T_0 = 66.25\, f_d^{-1}$, $n = 1.52$; 59th sensor: $\sigma = 9.33$ fTl/cm, $S_c(0) = 17761.77$ fT$^2 \cdot f_d^{-1}$/cm$^2$, $H_1 = 0.03$, $T_1 << T$, $T_0 << T$, $n = 1.41$.

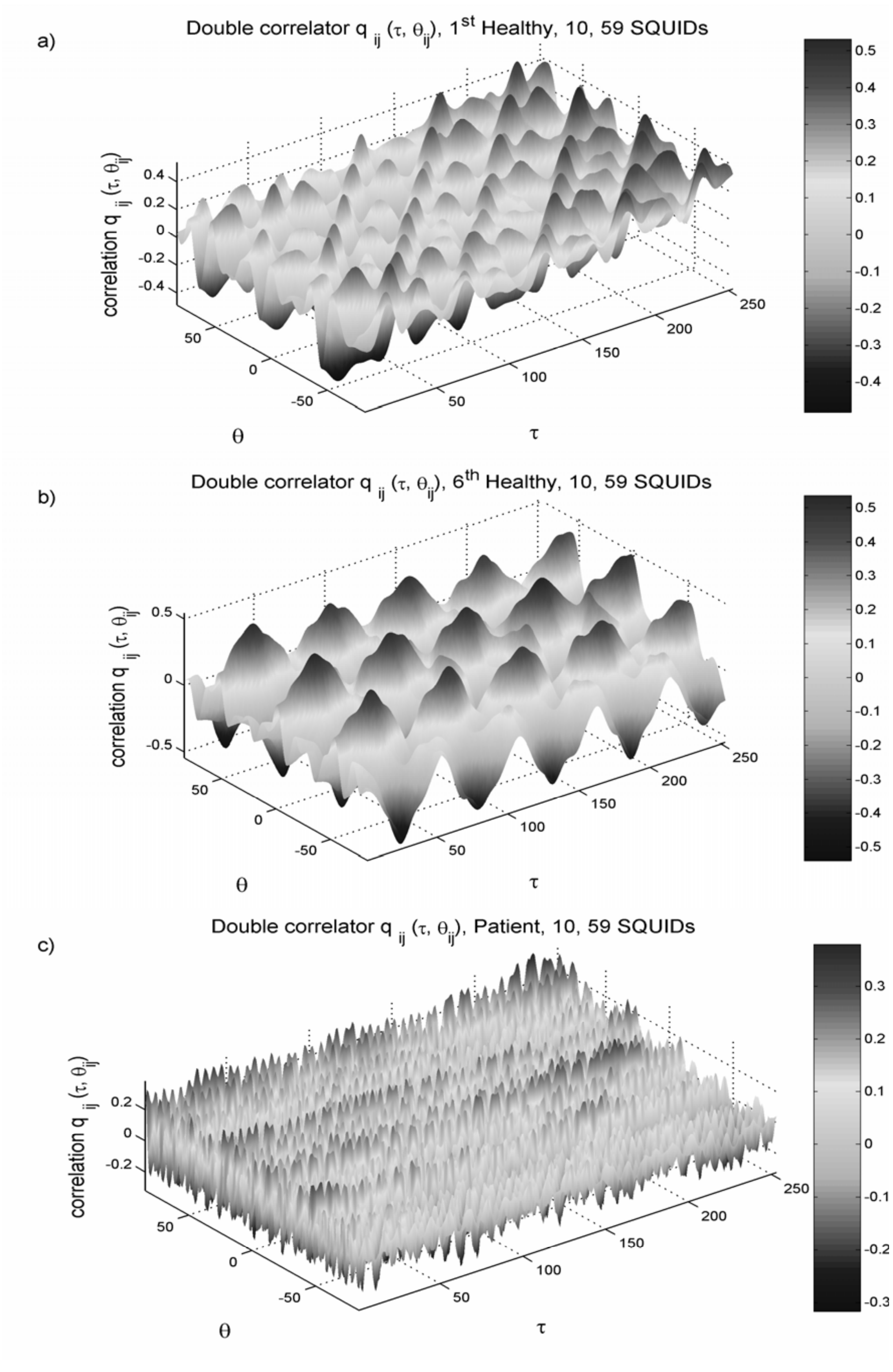

Fig. 7. Cross-correlations $q_{10\text{-}59}(\tau;\ \theta_{10\text{-}59})$ between the MEG signals at sensors 10 and 59 for the 1st control subject (a), 6th control subject (b), and PSE patient (c).

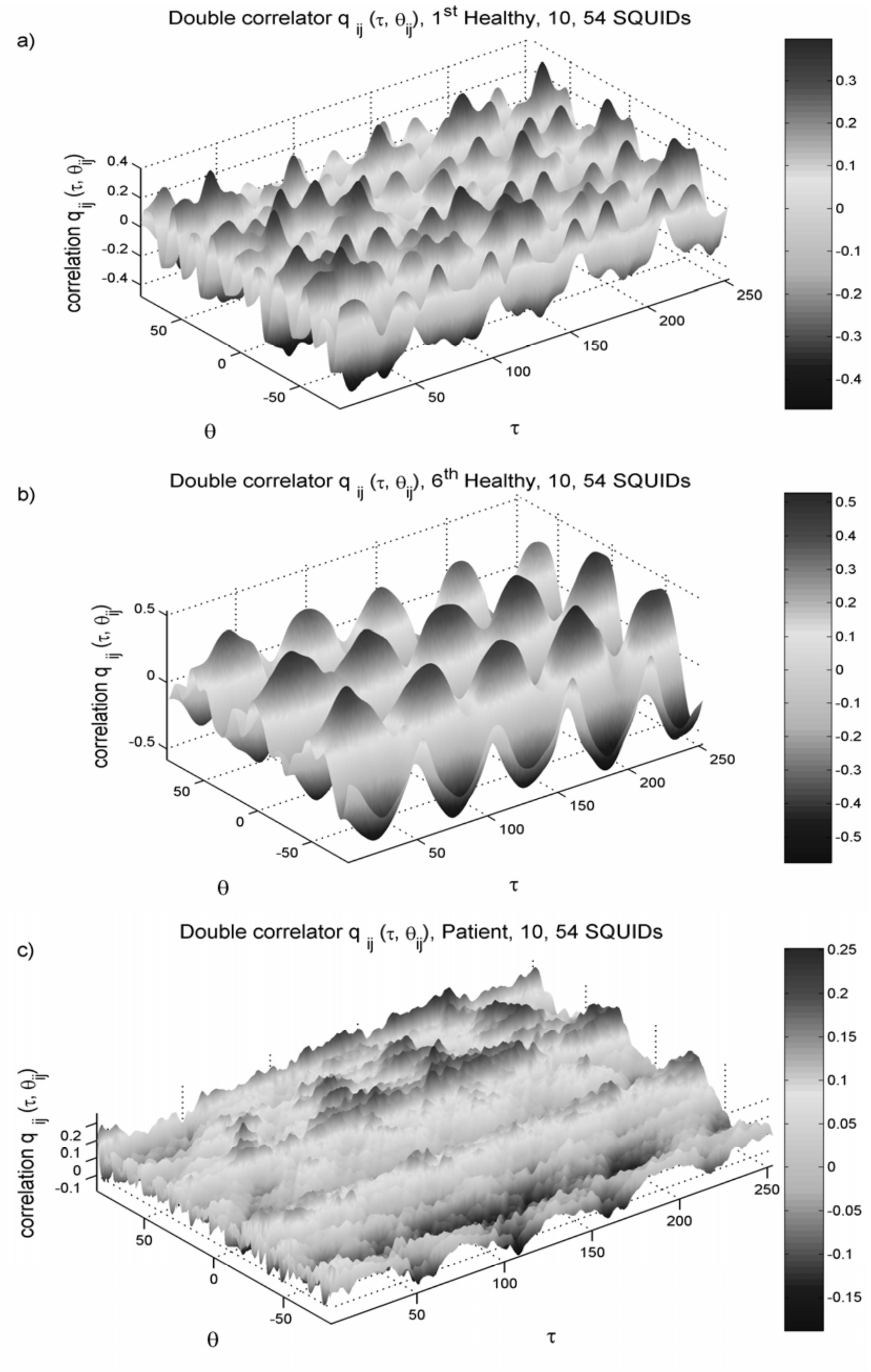

Fig. 8. Cross-correlations $q_{10\text{-}54}(\tau;\ \theta_{10\text{-}54})$ between the MEG signals at sensors 10 and 54 for the 1st control subject (a), 6th control subject (b), and PSE patient (c).

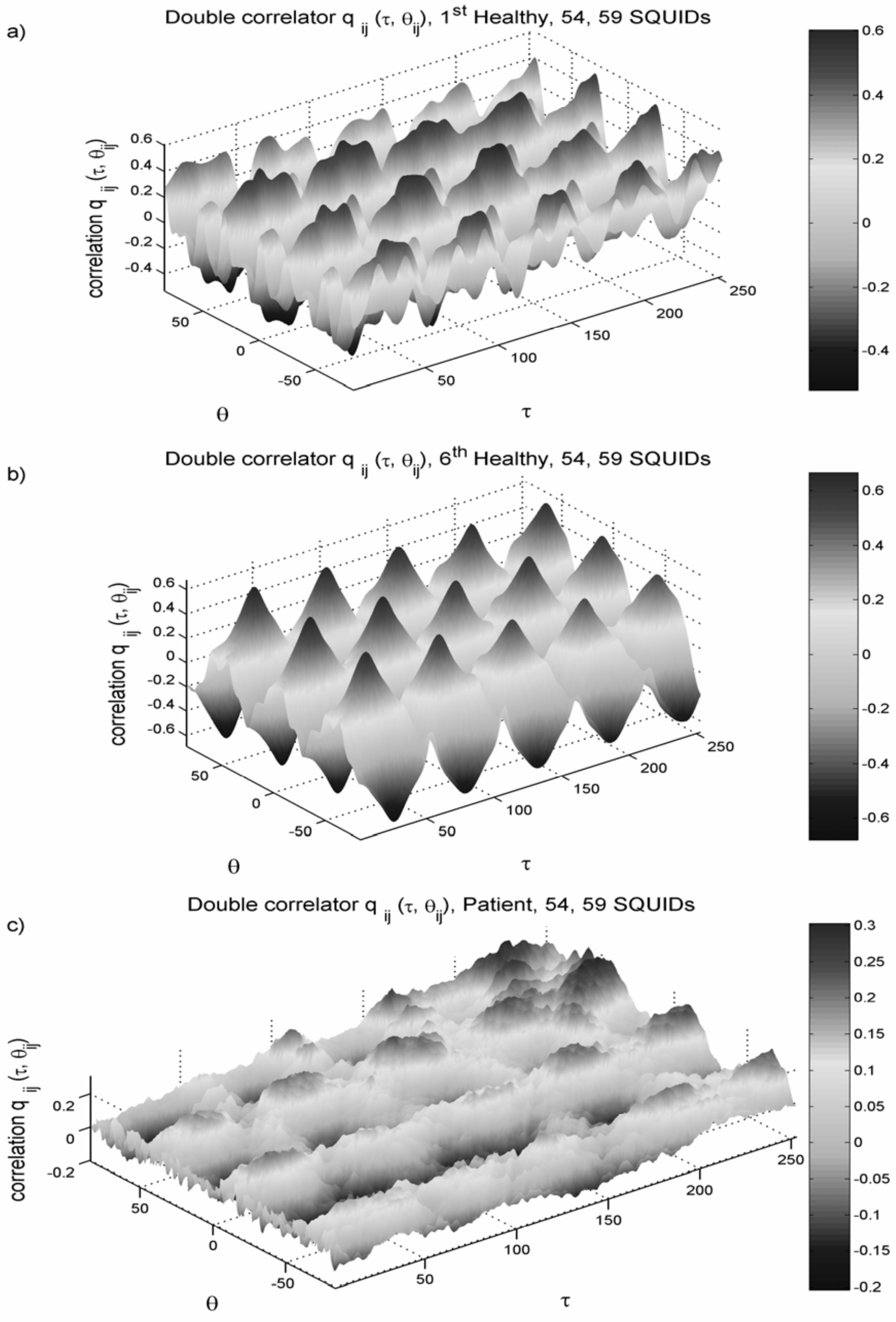

Fig. 9. Cross-correlations $q_{54\text{-}59}(\tau;\ \theta_{54\text{-}59})$ between the MEG signals at sensors 54 and 59 for the 1st control subject (a), 6th control subject (b), and PSE patient (c).

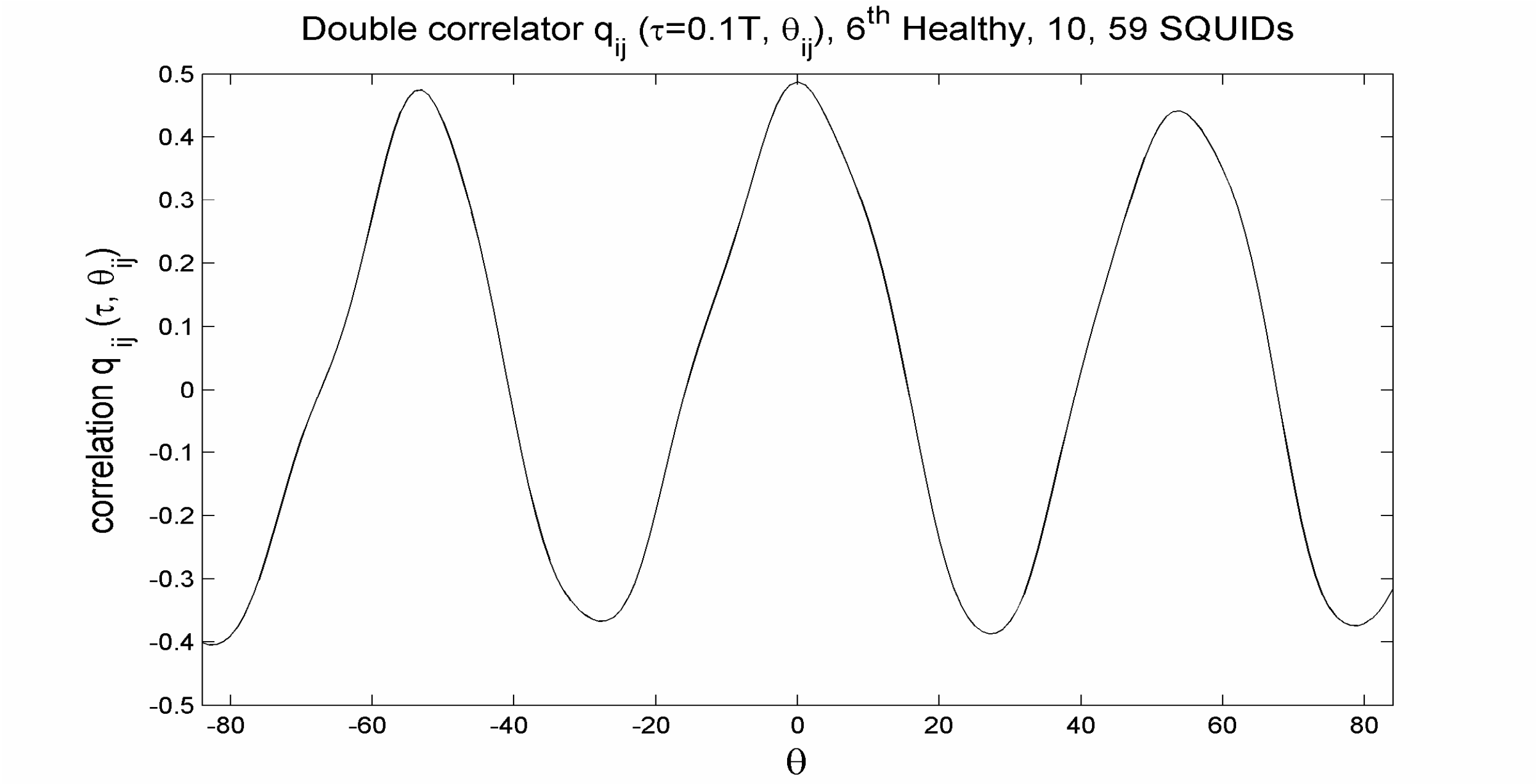

Fig. 10. Cross-section at $\tau^0 = 0.1\ T$ of the cross-correlation $q_{10\text{-}59}(\tau;\ \theta_{10\text{-}59})$ shown in Fig. 7b for the 6[th] control subject.

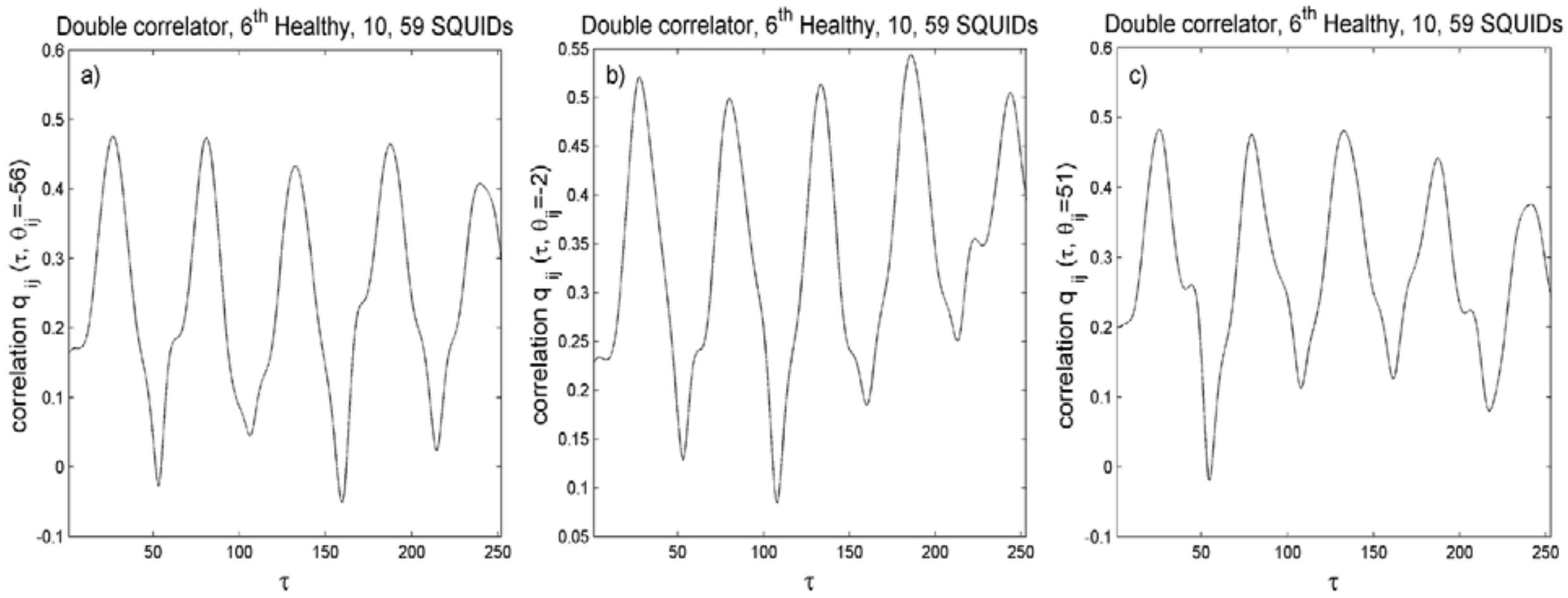

Fig. 11. Variations $q_{10\text{-}59}(\tau;\ \theta_{10\text{-}59}^{0+})$ at $\theta_{10\text{-}59}^{0+} = -56\, f_d^{-1}$ (a), $-2\, f_d^{-1}$ (b) and $51\, f_d^{-1}$ (c) for the 6[th] control subject.

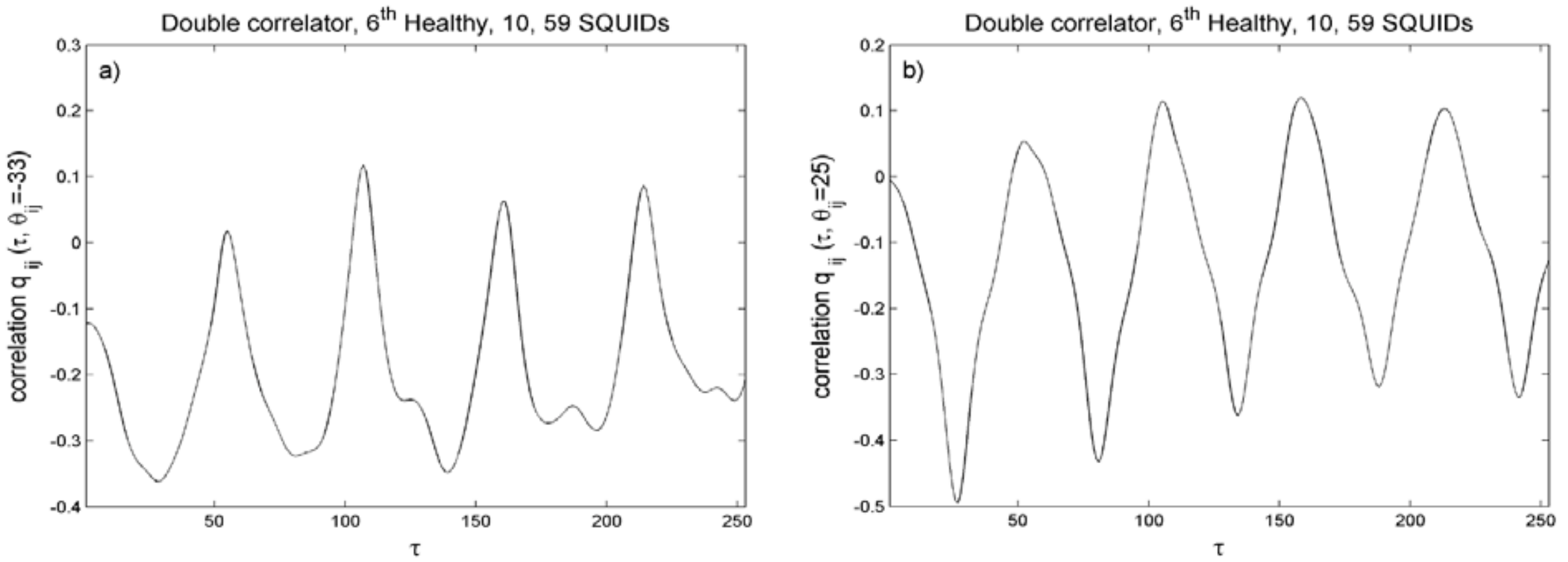

Fig. 12. Variations $q_{10\text{-}59}(\tau;\ \theta_{10\text{-}59}^{0-})$ at $\theta_{10\text{-}59}^{0-} = -33\, f_d^{-1}$ (a) and $25\, f_d^{-1}$(b) for the 6[th] control subject.

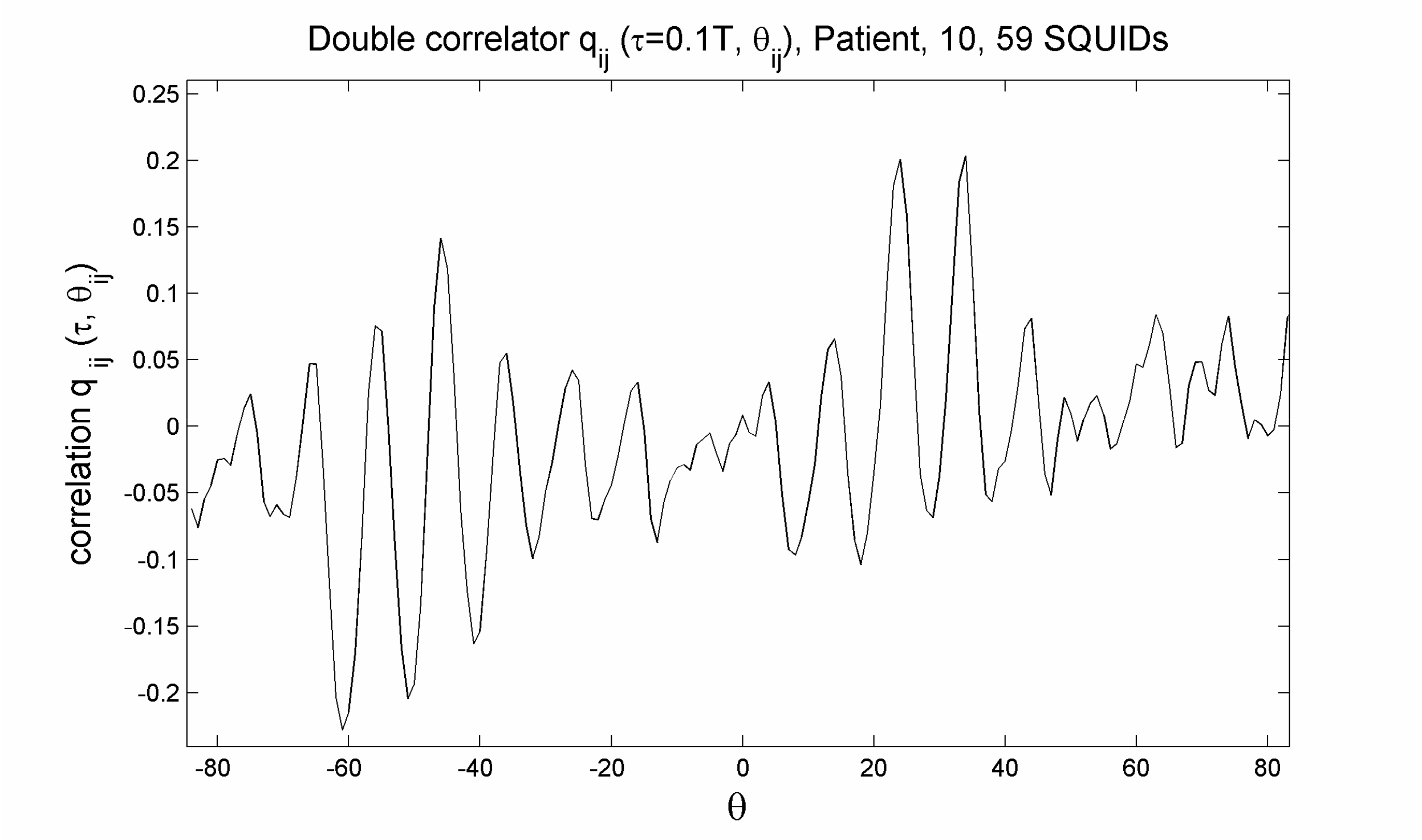

Fig. 13. Variation $q_{10\text{-}59}(84 f_d^{-1};\ \theta_{10\text{-}59})$ for the PSE patient.

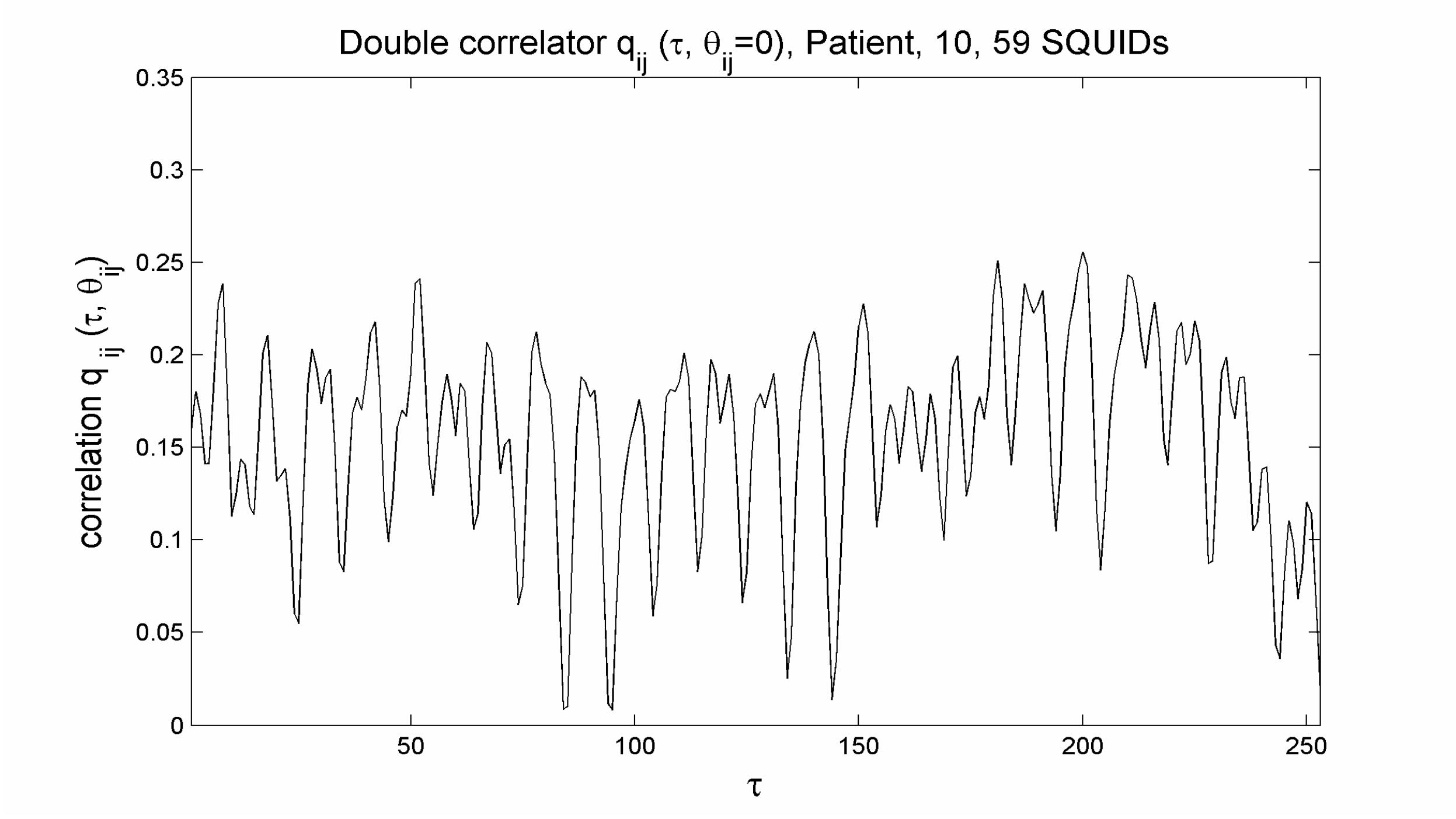

Fig. 14. Variation $q_{10\text{-}59}(\tau;\ 0)$ for the PSE patient.